\documentclass[]{article}
\usepackage[english]{babel}
\usepackage[utf8]{inputenc}
\usepackage{amsmath}
\usepackage{graphicx}
\usepackage[colorinlistoftodos]{todonotes}
\usepackage{authblk}
\usepackage{hyperref}
\usepackage{ulem}
\usepackage[margin=1in]{geometry}

\usepackage{natbib}
\setcitestyle{authoryear,open={(},close={)}}

\title{Challenges and Advances in Modeling of the Solar Atmosphere: \\A White Paper of Findings and Recommendations }

\author[1]{Serena Criscuoli}
\author[1,2]{Maria Kazachenko}
\author[3]{Irina Kitiashvili}
\author[3,4]{Alexander Kosovichev}
\author[1]{Valent\'in Mart\'inez Pillet}
\author[4]{Gelu Nita}
\author[5,6]{Viacheslav Sadykov}
\author[3]{Alan Wray}

\affil[1]{National Solar Observatory}
\affil[2]{University of Colorado, Boulder}
\affil[3]{NASA Ames Research Center}
\affil[4]{New Jersey Institute of Technology}
\affil[5]{Bay Area Environmental Research Institute}
\affil[6]{Georgia State University}

\date{December 30, 2020}
\begin{document}
\maketitle


\begin{abstract}
The next decade will be an exciting period for solar astrophysics, as new ground- and space-based instrumentation will provide unprecedented observations of the solar atmosphere and heliosphere. The synergy between modeling effort and comprehensive analysis of observations is crucial for the understanding of the physical processes behind the observed phenomena. However, the unprecedented wealth of data on one hand, and the complexity of the physical phenomena on the other, require the development of new approaches in both data analysis and numerical modeling. In this white paper, we summarize recent numerical achievements to reproduce structure, dynamics, and observed phenomena from the photosphere to the low corona and outline challenges we expect to face for the interpretation of future observations. 

\end{abstract}

\section{A forthcoming exciting Multi-Messenger era}
Investigations of physical processes occurring on the Sun, and their effects on the Earth, its atmosphere, and its space environment, exploit a large variety of messengers to provide the needed information. “Classical” photon-based astronomical observations create data such as disk-integrated spectra and full-disk and high spatial resolution images in wavelengths from radio to X-rays, as well as helioseismic imaging of the solar interior. The nearness of the Earth to the Sun allows measurements difficult or impossible for other stars,  such as in-situ measurements of solar wind properties (e.g., chemical composition, magnetic field, particle energy distributions, etc.), detection of neutrinos emitted in the solar core, and geomagnetic indices (which include direct measurements of properties of the magnetic field as well as indirect indices such as cosmogenic isotopes).  Each of these contributes to providing unique information about how energy is generated in the Sun, transported into the atmosphere and heliosphere, and  dissipated through a variety of phenomena on temporal scales of seconds to millennia.  

The next decade will be a new exciting period as new ground- and space-based instrumentation will provide unprecedented observations of the solar atmosphere and heliosphere. In particular, in 2021 the Daniel K. Inouye Solar Telescope {\citep[DKIST, ][]{rimmele2020}} will start to produce observations of the photosphere and chromosphere at extremely high spatial and temporal resolutions; these will be accompanied by spectral and spectropolarimetric observations of the corona at spectral ranges never or rarely observed before. The Parker Solar Probe (PSP, launched in August 2018) provides in-situ measurements of solar wind properties at unprecedented proximity to the Sun \citep{Fox2016}. The Solar Orbiter mission (launched in 2020) combines both in-situ and remote-sensing instrumentation and gives us the opportunity to expand our understanding about solar interior dynamics, the evolution and structure of polar magnetic fields, and the fast solar wind \citep{Muller2020}. Base knowledge from currently available observations from space missions (e.g., SDO, IRIS, HINODE, STEREO, THEMIS) and ground-based observatories (e.g., SOLIS, GST, CoMP, GONG, ALMA, EOVSA), accompanied by upcoming observations, will take us to a new level of understanding of solar dynamics and activity and their impact on the Heliospheric and Earth environments.

The synergy between an advanced modeling effort and a comprehensive analysis of observations has great potential to shed light on the physical processes behind the observed phenomena. However, as we approach smaller scales in models and observations, challenges emerge that require the development of new approaches in data analysis and numerical modeling. 

\section{Modeling of Solar Atmosphere Dynamics}

Substantial development of theoretical and numerical approaches, supported by continually growing computational capabilities and novel observing technologies, are changing our view of solar phenomena. Thanks to such developments, we have been slowly moving away from a simplistic, one-dimensional, static interpretation of solar observations toward models that take into account the multidimensional nature of the observed phenomena as well as the energy transport mechanisms occurring between different spatial and temporal scales. However, the complexity of the multi-scale dynamics of turbulent magnetized plasma in highly stratified inhomogeneous conditions prevents us from creating a single universal model to reproduce the solar dynamics. Therefore, despite the joint goal, modeling approaches are often targeted to a specific class of problems. For instance, realistic-type simulations are focused on capturing a wide range of the physical phenomena in the solar plasma (e.g., turbulence, radiation, ionization, magnetic fields, abundances) to reproduce observed processes and phenomena.
On the other hand, data-driven models attempt to reproduce a particular event or events by forcing a model solution toward specific observations to create an accurate physical interpretation of the observed dynamics.

Magnetohydrodynamic (MHD) simulations in the Sun-in-a-box regime aim to solve the equations of conservation  of mass, momentum, energy, and magnetic flux in a highly stratified compressible medium, utilizing 3D multi-group radiative energy transfer between the fluid elements, a real-gas equation of state, ionization and excitation of all abundant species, and magnetic effects. Because of their computational cost, the resulting models cover only a small area on the Sun with a spatial resolution of tens of kilometers or higher. Nevertheless, this approach demonstrates its capability to reproduce solar observations. In particular, `ab-initio' modeling of the solar magnetoconvection and atmosphere pioneered by \cite{Nordlund1989} demonstrated the importance of the effects of 3D turbulent dynamics in reproducing spectroscopic observations \citep{Asplund2009}.  Since then, such models have allowed us to make progress in understanding the near surface structuring and dynamics of solar plasma and magnetic fields. In particular,  they have provided insight about granulation \citep[e.g.,][]{Stein2001}, supersonic horizontal flows \citep[e.g.,][]{Vitas2011}, fine structuring of sunspot umbrae \citep[e.g.,][]{Schuessler2006} and penumbrae \citep[e.g.,][]{Kitiashvili2009,Rempel2009,Rempel2009a,SiuTapia2018}, convective collapse \citep[e.g.,][]{Skartlien2000,Hewitt2014}, explosive events \citep[e.g.,][]{Kitiashvili2013,Danilovic2016,Chatterjee2016}, small-scale dynamos and self-formed magnetic structures \citep[e.g.,][]{Vogler2007,Kitiashvili2010,Kitiashvili2015,Stein2012,Rempel2014}, the formation of magnetic structures from predefined conditions such as twisted flux ropes or specific magnetic field topologies \citep[e.g.,][]{Rempel2009,Cheung2010,Fang2015} and others.

\begin{figure}[t] 
    \centering
    \includegraphics[scale=0.4]{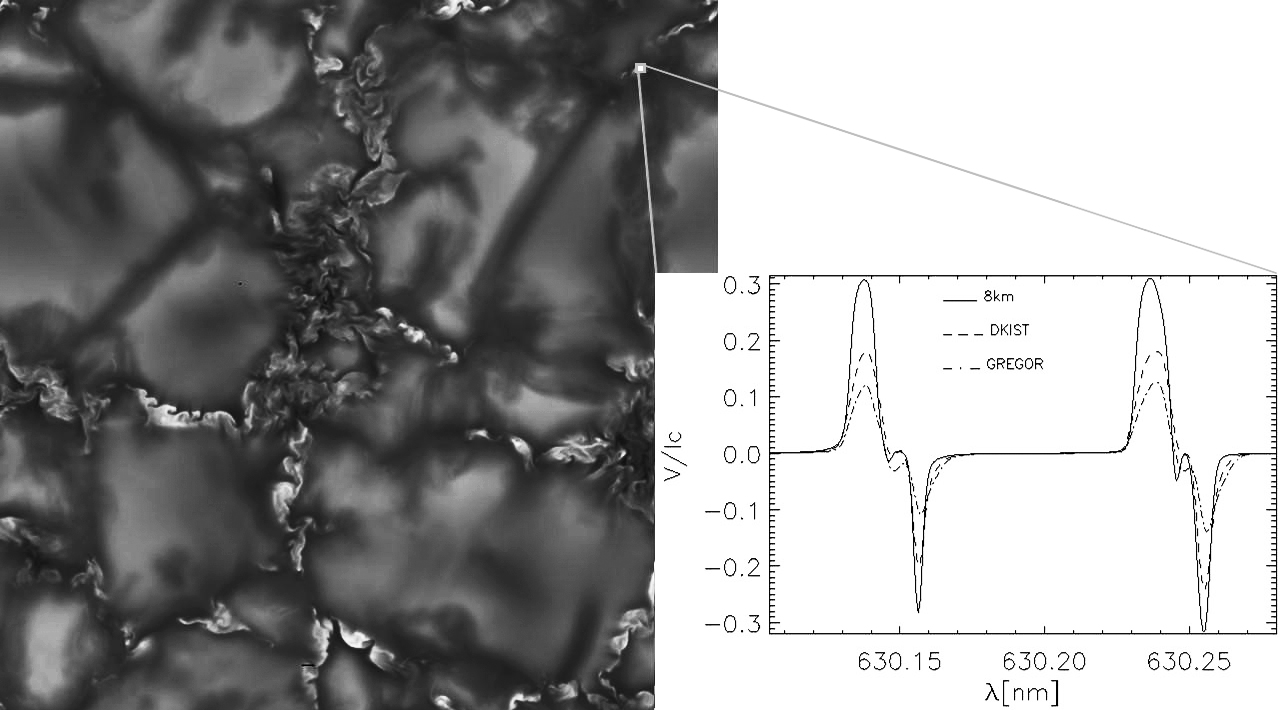}
    \caption{Synthetic continuum image obtained from a MuRAM MHD model \citep{Rempel2014} shows the the granular structure and magnetized bright structuring in the intergranular lanes. A probe of Stokes V spectra (Fe I 630~nm) illustrates the dependence of the line profiles on the spatial resolution:  full numerical resolution(8~km, solid curve), corresponding resolutions of DKIST (40~km, dashed curve) and the GREGOR telescope (80~km, dashed-dotted curve).
}
    \label{fig:stokesV}
\end{figure}

Thanks to intense cross-validation of the numerical models with observations, which has allowed achieving a high-degree realism of the simulated solar plasma, these 3D radiative models have become capable of providing critical support in the interpretation of spectroscopic and spectropolarimetric observations from the ground and in space. Figure~\ref{fig:stokesV} shows an example of continuum intensity synthesized from an MHD model \citep{Rempel2014} obtained with the MuRAM code \citep{Vogler2005} and a comparison of the Stokes V profiles obtained for the numerical resolution and resolutions corresponding to 4-m DKIST and 1.5-m GREGOR observations. Discrepancies in resulting line profiles illustrate how the spatial resolution can affect the accuracy of magnetic field measurements. Analysis of such discrepancies helps us to estimate uncertainties in the inferred physical and dynamical properties of the plasma. Indeed, the leveraging of cross-comparison between observations and simulations is crucial for several scientific issues discussed in the DKIST Critical Science Plan \citep{Rast2020}.

Coronal magnetic fields play a key role in determining when and where solar flares and coronal mass ejections (CMEs) occur \citep{Gibson2020}. Remote sensing of the coronal magnetic fields has not been possible in the past, thus only reconstructions based on magnetic field measurements at the photosphere/chromosphere level could be employed to model the 3D structure of the coronal magnetic field. However, newly developed observational techniques, such as visible/infrared (VIR) coronal seismology \citep{Yang2020a, Yang2020b} and microwave interferometry \citep{Fleishman2020, Chen2020}, have demonstrated the feasibility of off-limb and on-limb observations to be routinely performed in the near future using a new generation of already developed instruments, such as DKIST and The Expanded Owens Valley Solar Array \citep[EOVSA, ][]{EOVSA}, and proposed  instruments, such as the full-Sun Coronal Solar Magnetism Observatory \citep[COSMO, ][]{Tomczyk} and the Frequency Agile Solar Radiotelescope \citep[FASR, ][]{FASR}

Nevertheless, given the remote sensing nature of these newly available observations, they are generally limited by a line-of-sight ambiguity, which can only be resolved by combining them with sophisticated, yet feasible, multi-wavelength forward-fitting techniques and data-constrained modeling frameworks, such those offered by community-developed and maintained modeling tools like FORWARD \citep{Gibson2016} and GX Simulator \citep{Nita2015, Nita2018}.

Modeling of the magnetic field in the corona is essential for the study of the origin of  the solar active phenomena. There are primarily two groups of models to simulate evolving coronal magnetic fields in 3D: quasi-static (or time-independent) and dynamic (time-dependent). In the quasi-static group, potential, linear, and non-linear force-free field (NLFFF) extrapolations have been developed. These models apply the vacuum-limit assumption, which reasonably assumes that magnetic pressure dominates the gas pressure (low-beta regime). Although this modeling framework assumes that the corona is in a static force balance that is not affected by states at earlier times,  \citet{Fleishman2018} have shown that evolving sequences of such modeling snapshots may be used to infer the evolution of eruptive phenomena.

Another, more physics-based group of models are time-dependent models. In this group,  ``data-driven'' (DD) simulations, i.e., simulations where successive photospheric magnetic field observations are used to evolve the model's photospheric field, show great promise for investigating the 3D structure of the coronal magnetic field. Lately, two types of DD models have been developed: magnetofrictional \citep[e.g.][]{Cheung2012} and MHD models \citep[e.g.][]{Hayashi2018}. The magnetofrictional model assumes that the plasma velocity in the MHD induction equation is proportional to the local Lorentz force, leading to a relaxation of a magnetic configuration toward a force-free state. It is more computationally efficient than MHD and is suitable for the description of the slow quiescent evolution of active regions, but not for flares. The MHD models explicitly solve a full set of MHD equations including the plasma properties. The MHD approach is suitable for rapid evolution during flares but is too computationally expensive to model the long-term quiescent evolution of active regions. A hybrid framework, where the MF model is used to model quiescent periods of AR evolution and the MHD model is used to model flaring periods of AR evolution, has been recently developed within the Coronal Global Evolutionary model \citep{Hoeksema2020}.

  The major difference between data-driven and data-inspired models are the observations that the data-driven models make use of as lower boundary conditions. These are typically some combination of magnetic, velocity, and electric fields. Recently several approaches have been developed to derive electric fields using observations of the vector magnetic fields and Doppler velocities in the photosphere \citep[see e.g.][]{Fisher2020}.


The current effort to capture solar dynamics from the convection zone to the corona faces the challenge of numerically describing atmospheric layers with dramatic variations of plasma conditions. 
In addition, a Non-Local Thermodynamic Equilibrium (NLTE) approach is necessary on one hand to interpret chromospheric and transition region diagnostics 
 (e.g., H-alpha, UV, IR Ca\,II\,lines, Mg\,II\,h\&k, He\,I, Ly-alpha)  \citep[e.g.][]{Leenaarts2013a, Stepan2015} and therefore derive reliable observational constraints, and, on the other hand, to properly describe the interaction of radiation and matter in the higher layers of the atmosphere.
 One of the most advanced codes in this respect is \textit {Bifrost} \citep{Gudiksen2011}, which,  although in a simplified manner,  includes NLTE processes in the energy and particle balance \citep{Carlsson2016,Hansteen2017}. Significant improvements have been further achieved by including additional effects, such as the Hall effect and ambipolar diffusion \citep[e.g.,][]{Martinez-Sykora2012, Khomenko2020}, and the development of approximations to describe chromospheric radiative cooling and heating \citep{Carlsson2012}.

\begin{figure}[t]
    \centering
    \includegraphics[width=0.75\linewidth]{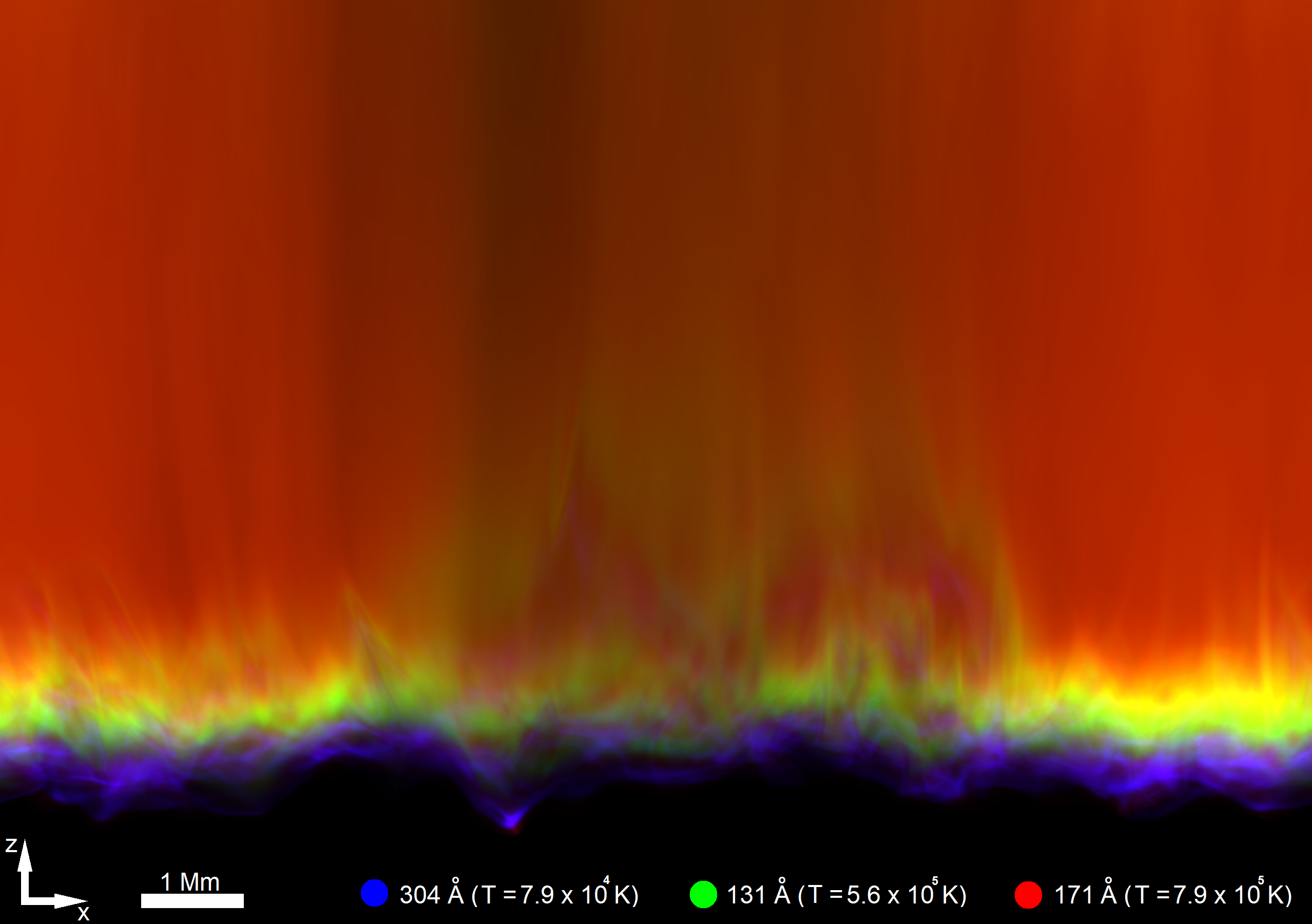}
    \caption{Composite illustration of the Solar Dynamics Observatory Atmospheric Imaging Assembly (SDO/AIA) synthetic emission for a side view of a quiet Sun region, simulated with the StellarBox code \citep{Wray2018}. Red color corresponds to the synthetic SDO/AIA 171\,\AA{} emission (T$=7.9\times{}10^{5}$\,K), green corresponds to 131\,\AA{} (T$=5.6\times{}10^{5}$\,K), blue corresponds to 304\,\AA{} (T$=7.9\times{}10^{4}$\,K). The domain has a horizontal size of 12.8\,Mm and includes 10\,Mm of the solar atmosphere from the surface to corona.}
    \label{fig:aiacorona}
\end{figure}

However, the computational cost of NLTE calculations poses significant limitations on the size of the computational domain and spatial resolution.
Therefore, modeling of the solar dynamics in broad regimes, from the interior to the corona, necessarily relies on the LTE approximation.  \citep[e.g.,][]{Rempel2017,Wray2018}. 
 In spite of this simplification, various codes have been used to simulate the domain from the photosphere to the corona, allowing the study of phenomena naturally driven by near-surface plasma dynamics, and addressing issues as coronal heating, the onset of flares, and others  \citep[e.g.,][]{Gudiksen2005,Carlsson2012,Rempel2017,Cheung2019}. These models provide theoretical support for the interpretation of observations. An example of synthetic SDO/AIA observations of the limb view of the modeled solar corona~\citep{Kitiashvili2020} is presented in Figure~\ref{fig:aiacorona}. 

 As alluded to above, the use of state-of-the-art atmosphere models to interpret solar observations is further complicated by difficulties inherent to the synthesis of specific observables.  
Several radiative transfer codes have been developed for this purpose, each with certain advantages and restrictions. For instance, the Spinor \citep{Frutiger2000} code is relatively fast, but computations are restricted by the Local Thermodynamic Equilibrium (LTE) approximation. The NICOLE code \citep{SocasNavarro2015} enables calculations in both LTE and NLTE, but the implementation of the NLTE approximation does not include partial frequency redistribution (PRD) effects, which are important for the modeling of chromospheric lines, such as the Ca II H \& K, and Mg II UV lines. The RH code \citep{Pereira2015} is also able to perform calculations in both approximations with a primary focus on the accurate implementation of NLTE effects in different types of geometries. The Multi3D code \citep{Leenaarts2009} is developed for massive calculations from 3D models of the solar atmosphere of a wide range of spectral lines using the NLTE approximation. However, the inclusion of detailed multi-dimensional radiative transfer effects comes at the cost of the increased computational time.  Moreover, the description of these physical effects is numerically challenging, especially in the chromosphere where large spatial and temporal gradients often make the codes unstable. Ultimately, the choice of the particular radiative transfer code depends on the task and the related approximations most suitable for the physical processes one wants to study.

The interpretation of observational results often relies on solving the inversion problem, i.e., on modeling the physical characteristics of the atmosphere which produced the observed emission and spectra. 
Different codes are available to perform spectral ad spectropolarimetric inversions of observed spectra, but, similarly to the codes available for forward modeling, the vast majority are targeted at the interpretation of specific types of observations. For instance, codes developed for the interpretation of lines forming in the photosphere and chromosphere are often developed by integrating the radiative transfer codes described in the previous section with minimization algorithms that optimize the physical and magnetic stratification of the atmosphere (examples include the widely used NICOLE and SPINOR and the newly developed STiC and DeSIRE, based on the RH code). As such, these inversion codes suffer of the same limitations described above. Other codes have been or are under development to interpret specific observations, as for instance observations in the He I  1083.0 nm line (Hazel), limb observations in chromospheric lines (NICOLE), and flares (STiC). Interpretation of coronal observations is instead performed by using very different approaches, based on differential emission measures of the coronal plasma from the related EUV observations \citep[e.g.,][]{Cheung2015}. One of the major drawbacks of most of inversion techniques is their extremely high computational cost.  
One solution is to represent the highly-nonlinear mapping of the observed emission and line profiles to the structure of the atmosphere by a neural network trained on the set of physics-based inversions. Such an approach has been successfully applied for fast inversion of IRIS data \citep{SainzDalda2019} and SDO/AIA EUV emission \citep{Wright2019}. Although uncertainty quantification for such machine-learning-based models is challenging and still open, this is currently the only way to provide timely data products under one pipeline with the observational data. The development of such models and understanding and quantifying the related uncertainties should be encouraged.

\section{Access to computational resources and model data sharing}
 High spatial and temporal resolutions, augmented by the spectropolarimetric dimensions, dramatically increase the data-volume of both observational and synthetic data. Such massive data flows significantly limit the dissemination of these data among the community. 

On the one hand, sharing of the modeling and synthetic-observables data sets requires dramatic storage resources. On the other hand, taking these data and analyzing them on a local workstation is often impossible. As a result, the modeling data are highly truncated (e.g., include only selected layers, model parameters, and/or limited time-series). 
The problem is becoming more severe with time to the point that web links to the data are sometimes failing. Therefore, the development of a data-sharing platform and data-request system will benefit both potential customers and data providers. 

 Of particular importance is the collection of synthetic observables obtained with different radiative transfer codes and MHD models. 
 A database of synthetic observables, obtained by combining state-of-the-art model atmospheres and radiative transfer computations, will support the scientific community in several crucial ways, such as the interpretation of observations, the development and validation of new data analysis techniques, and the validation of new instrumentation and data reduction pipelines. Furthermore, a wider dissemination of synthetic data will improve our understanding of model limitations, thus fostering the development of new numerical techniques and a better description of the physical processes that drive the evolution of the solar plasma and magnetic fields.

\section{Recommendations}

 The development of numerical models and tools is essential for a robust physical interpretation of the observational data acquired by the DKIST and other ground and space missions that have recently come online. The analysis of spectral and spectropolarimetric observations acquired at high spatial and temporal resolution and at high spectropolarimetric sensitivity in different spectral ranges, requires the development of models capable of describing the non-linear connection between the local physical and dynamical properties of plasma in a radiating, turbulent, and magnetic environment. Despite significant advances in modeling solar dynamics, the cross-validation of models and observations is essential to detect any missed components of the observed phenomena.

Thus it is crucial to support the multi-level integration of modeling efforts and observations. In particular, we identify the following critical needs:
\begin{itemize}
    \item Improve the development of high-fidelity 3D MHD codes by 1) removing restrictions on code development commitments in proposal AOs, and 2) establishing new funding opportunities for the development of codes and data analysis approaches, including machine learning, data assimilation, and others.
    \item Encourage cross-comparison of observations and simulations and support the development of new tools for efficient analysis of big multi-dimensional data sets.
    \item Increase investment in high-end-computing (HEC) facilities in terms of both computing nodes and storage. The storage and computational requirements are amplified by including more physical effects into the models and for the post-processing of the resulting data. Improvement in HEC capabilities will increase workflow performance and will significantly speed up development of predictive capabilities for both global solar variability and short-term, high-energy release activity such as flares, coronal mass ejections, etc. Similarly, the management and analysis of large observational datasets (for instance through spectropolarimetric inversions) requires increasing computational and storage resources.
    \item Allow trial allocations to demonstrate the feasibility of proposed projects and to speed up the boarding process when the first full allocation is awarded. Trial allocations will provide quick access to computational resources when needed and will help foster a new generation of scientists and developers.
    \item Establish a NASA-NSF HEC collaboration to enable efficient workflow. Code performance can depend on supercomputer architecture, infrastructure, and available resources. Consequently, rigid funding linked to a specific supercomputer system can impede project success.
    \item Support the development of open access policies for models and synthetic data, in a manner similar to policies implemented for observational data acquired by NASA and NSF facilities.
    \item Support the development of platforms for requesting synthetic data targeted at the interpretation of specific observations. 
    \item Support the development of platforms for the distribution of models and synthetic data to leverage existing capabilities and efforts, to improve collaboration efficiency, and to stimulate interdisciplinary integration.

\end{itemize}

\section*{Acknowledgments}
This material is based upon work supported by the National Science Foundation under Grant AGS-1743321. Any opinions, ﬁndings, and conclusions or recommendations expressed in this material are those of the authors and do not necessarily reﬂect the views of the National Science Foundation.The National Solar Observatory is operated by the Association of Universities for Research in Astronomy, Inc. (AURA)
under cooperative agreement with the National Science Foundation. 
\bibliography{ref.bib}

\begin{thebibliography}{62}
\providecommand{\natexlab}[1]{#1}
\providecommand{\url}[1]{\texttt{#1}}
\expandafter\ifx\csname urlstyle\endcsname\relax
  \providecommand{\doi}[1]{doi: #1}\else
  \providecommand{\doi}{doi: \begingroup \urlstyle{rm}\Url}\fi

\bibitem[{Asplund} et~al.(2009){Asplund}, {Grevesse}, {Sauval}, and
  {Scott}]{Asplund2009}
M.~{Asplund}, N.~{Grevesse}, A.~J. {Sauval}, and P.~{Scott}.
\newblock {The Chemical Composition of the Sun}.
\newblock \emph{\araa}, 47\penalty0 (1):\penalty0 481--522, Sept. 2009.
\newblock \doi{10.1146/annurev.astro.46.060407.145222}.

\bibitem[{Bastian}(2003)]{FASR}
T.~S. {Bastian}.
\newblock {The Frequency Agile Solar Radiotelescope}.
\newblock \emph{Advances in Space Research}, 32\penalty0 (12):\penalty0
  2705--2714, Jan. 2003.
\newblock \doi{10.1016/S0273-1177(03)00903-7}.

\bibitem[{Carlsson} and {Leenaarts}(2012)]{Carlsson2012}
M.~{Carlsson} and J.~{Leenaarts}.
\newblock {Approximations for radiative cooling and heating in the solar
  chromosphere}.
\newblock \emph{\aap}, 539:\penalty0 A39, Mar. 2012.
\newblock \doi{10.1051/0004-6361/201118366}.

\bibitem[{Carlsson} et~al.(2016){Carlsson}, {Hansteen}, {Gudiksen},
  {Leenaarts}, and {De Pontieu}]{Carlsson2016}
M.~{Carlsson}, V.~H. {Hansteen}, B.~V. {Gudiksen}, J.~{Leenaarts}, and B.~{De
  Pontieu}.
\newblock {A publicly available simulation of an enhanced network region of the
  Sun}.
\newblock \emph{\aap}, 585:\penalty0 A4, Jan. 2016.
\newblock \doi{10.1051/0004-6361/201527226}.

\bibitem[{Chatterjee} et~al.(2016){Chatterjee}, {Hansteen}, and
  {Carlsson}]{Chatterjee2016}
P.~{Chatterjee}, V.~{Hansteen}, and M.~{Carlsson}.
\newblock {Modeling Repeatedly Flaring {\ensuremath{\delta}} Sunspots}.
\newblock \emph{\prl}, 116\penalty0 (10):\penalty0 101101, Mar. 2016.
\newblock \doi{10.1103/PhysRevLett.116.101101}.

\bibitem[{Chen} et~al.(2020){Chen}, {Shen}, {Gary}, {Reeves}, {Fleishman},
  {Yu}, {Guo}, {Krucker}, {Lin}, {Nita}, and {Kong}]{Chen2020}
B.~{Chen}, C.~{Shen}, D.~E. {Gary}, K.~K. {Reeves}, G.~D. {Fleishman}, S.~{Yu},
  F.~{Guo}, S.~{Krucker}, J.~{Lin}, G.~M. {Nita}, and X.~{Kong}.
\newblock {Measurement of magnetic field and relativistic electrons along a
  solar flare current sheet}.
\newblock \emph{Nature Astronomy}, July 2020.
\newblock \doi{10.1038/s41550-020-1147-7}.

\bibitem[{Cheung} and {DeRosa}(2012)]{Cheung2012}
M.~C.~M. {Cheung} and M.~L. {DeRosa}.
\newblock {A Method for Data-driven Simulations of Evolving Solar Active
  Regions}.
\newblock \emph{\apj}, 757\penalty0 (2):\penalty0 147, Oct. 2012.
\newblock \doi{10.1088/0004-637X/757/2/147}.

\bibitem[{Cheung} et~al.(2010){Cheung}, {Rempel}, {Title}, and
  {Sch{\"u}ssler}]{Cheung2010}
M.~C.~M. {Cheung}, M.~{Rempel}, A.~M. {Title}, and M.~{Sch{\"u}ssler}.
\newblock {Simulation of the Formation of a Solar Active Region}.
\newblock \emph{\apj}, 720\penalty0 (1):\penalty0 233--244, Sept. 2010.
\newblock \doi{10.1088/0004-637X/720/1/233}.

\bibitem[{Cheung} et~al.(2015){Cheung}, {Boerner}, {Schrijver}, {Testa},
  {Chen}, {Peter}, and {Malanushenko}]{Cheung2015}
M.~C.~M. {Cheung}, P.~{Boerner}, C.~J. {Schrijver}, P.~{Testa}, F.~{Chen},
  H.~{Peter}, and A.~{Malanushenko}.
\newblock {Thermal Diagnostics with the Atmospheric Imaging Assembly on board
  the Solar Dynamics Observatory: A Validated Method for Differential Emission
  Measure Inversions}.
\newblock \emph{\apj}, 807\penalty0 (2):\penalty0 143, July 2015.
\newblock \doi{10.1088/0004-637X/807/2/143}.

\bibitem[{Cheung} et~al.(2019){Cheung}, {Rempel}, {Chintzoglou}, {Chen},
  {Testa}, {Mart{\'\i}nez-Sykora}, {Sainz Dalda}, {DeRosa}, {Malanushenko},
  {Hansteen}, {De Pontieu}, {Carlsson}, {Gudiksen}, and {McIntosh}]{Cheung2019}
M.~C.~M. {Cheung}, M.~{Rempel}, G.~{Chintzoglou}, F.~{Chen}, P.~{Testa},
  J.~{Mart{\'\i}nez-Sykora}, A.~{Sainz Dalda}, M.~L. {DeRosa},
  A.~{Malanushenko}, V.~{Hansteen}, B.~{De Pontieu}, M.~{Carlsson},
  B.~{Gudiksen}, and S.~W. {McIntosh}.
\newblock {A comprehensive three-dimensional radiative magnetohydrodynamic
  simulation of a solar flare}.
\newblock \emph{Nature Astronomy}, 3:\penalty0 160--166, Nov. 2019.
\newblock \doi{10.1038/s41550-018-0629-3}.

\bibitem[{Danilovic} et~al.(2016){Danilovic}, {van Noort}, and
  {Rempel}]{Danilovic2016}
S.~{Danilovic}, M.~{van Noort}, and M.~{Rempel}.
\newblock {Internetwork magnetic field as revealed by two-dimensional
  inversions}.
\newblock \emph{\aap}, 593:\penalty0 A93, Sept. 2016.
\newblock \doi{10.1051/0004-6361/201527842}.

\bibitem[{Fang} and {Fan}(2015)]{Fang2015}
F.~{Fang} and Y.~{Fan}.
\newblock {{\ensuremath{\delta}}-Sunspot Formation in Simulation of
  Active-region-scale Flux Emergence}.
\newblock \emph{\apj}, 806\penalty0 (1):\penalty0 79, June 2015.
\newblock \doi{10.1088/0004-637X/806/1/79}.

\bibitem[{Fisher} et~al.(2020){Fisher}, {Kazachenko}, {Welsch}, {Sun}, {Lumme},
  {Bercik}, {DeRosa}, and {Cheung}]{Fisher2020}
G.~H. {Fisher}, M.~D. {Kazachenko}, B.~T. {Welsch}, X.~{Sun}, E.~{Lumme}, D.~J.
  {Bercik}, M.~L. {DeRosa}, and M.~C.~M. {Cheung}.
\newblock {The PDFI\_SS Electric Field Inversion Software}.
\newblock \emph{\apjs}, 248\penalty0 (1):\penalty0 2, May 2020.
\newblock \doi{10.3847/1538-4365/ab8303}.

\bibitem[Fleishman et~al.(2018)Fleishman, Nita, Kuroda, Jia, Tong, Wen, and
  Zhizhuo]{Fleishman2018}
G.~Fleishman, G.~Nita, N.~Kuroda, S.~Jia, K.~Tong, R.~Wen, and Z.~Zhizhuo.
\newblock Revealing the evolution of non-thermal electrons in solar flares
  using 3d modeling.
\newblock \emph{Astrophysical Journal}, 859\penalty0 (1), 2018.

\bibitem[{Fleishman} et~al.(2020){Fleishman}, {Gary}, {Chen}, {Kuroda}, {Yu},
  and {Nita}]{Fleishman2020}
G.~D. {Fleishman}, D.~E. {Gary}, B.~{Chen}, N.~{Kuroda}, S.~{Yu}, and G.~M.
  {Nita}.
\newblock {Decay of the coronal magnetic field can release sufficient energy to
  power a solar flare}.
\newblock \emph{Science}, 367\penalty0 (6475):\penalty0 278--280, Jan. 2020.
\newblock \doi{10.1126/science.aax6874}.

\bibitem[{Fox} et~al.(2016){Fox}, {Velli}, {Bale}, {Decker}, {Driesman},
  {Howard}, {Kasper}, {Kinnison}, {Kusterer}, {Lario}, {Lockwood}, {McComas},
  {Raouafi}, and {Szabo}]{Fox2016}
N.~J. {Fox}, M.~C. {Velli}, S.~D. {Bale}, R.~{Decker}, A.~{Driesman}, R.~A.
  {Howard}, J.~C. {Kasper}, J.~{Kinnison}, M.~{Kusterer}, D.~{Lario}, M.~K.
  {Lockwood}, D.~J. {McComas}, N.~E. {Raouafi}, and A.~{Szabo}.
\newblock {The Solar Probe Plus Mission: Humanity's First Visit to Our Star}.
\newblock \emph{\ssr}, 204\penalty0 (1-4):\penalty0 7--48, Dec. 2016.
\newblock \doi{10.1007/s11214-015-0211-6}.

\bibitem[{Frutiger} et~al.(2000){Frutiger}, {Solanki}, {Fligge}, and
  {Bruls}]{Frutiger2000}
C.~{Frutiger}, S.~K. {Solanki}, M.~{Fligge}, and J.~H.~M.~J. {Bruls}.
\newblock {Properties of the solar granulation obtained from the inversion of
  low spatial resolution spectra}.
\newblock \emph{\aap}, 358:\penalty0 1109--1121, June 2000.

\bibitem[{Gary} et~al.(2018){Gary}, {Chen}, {Dennis}, {Fleishman}, {Hurford},
  {Krucker}, {McTiernan}, {Nita}, {Shih}, {White}, and {Yu}]{EOVSA}
D.~E. {Gary}, B.~{Chen}, B.~R. {Dennis}, G.~D. {Fleishman}, G.~J. {Hurford},
  S.~{Krucker}, J.~M. {McTiernan}, G.~M. {Nita}, A.~Y. {Shih}, S.~M. {White},
  and S.~{Yu}.
\newblock {Microwave and Hard X-Ray Observations of the 2017 September 10 Solar
  Limb Flare}.
\newblock \emph{\apj}, 863\penalty0 (1):\penalty0 83, Aug 2018.
\newblock \doi{10.3847/1538-4357/aad0ef}.

\bibitem[{Gibson} et~al.(2016){Gibson}, {Kucera}, {White}, {Dove}, {Fan},
  {Forland}, {Rachmeler}, {Downs}, and {Reeves}]{Gibson2016}
S.~{Gibson}, T.~{Kucera}, S.~{White}, J.~{Dove}, Y.~{Fan}, B.~{Forland},
  L.~{Rachmeler}, C.~{Downs}, and K.~{Reeves}.
\newblock {FORWARD: A toolset for multiwavelength coronal magnetometry}.
\newblock \emph{Frontiers in Astronomy and Space Sciences}, 3:\penalty0 8, Mar.
  2016.
\newblock \doi{10.3389/fspas.2016.00008}.

\bibitem[Gibson et~al.(2020)Gibson, Malanushenko, de~Toma, Tomczyk, Reeves,
  Tian, Yang, Chen, Fleishman, Gary, Nita, Pillet, White, Bak-Steslicka,
  Dalmasse, Kucera, Rachmeler, Raouafi, and Zhao]{Gibson2020}
S.~E. Gibson, A.~Malanushenko, G.~de~Toma, S.~Tomczyk, K.~Reeves, H.~Tian,
  Z.~Yang, B.~Chen, G.~Fleishman, D.~Gary, G.~Nita, V.~M. Pillet, S.~White,
  U.~Bak-Steslicka, K.~Dalmasse, T.~Kucera, L.~A. Rachmeler, N.~E. Raouafi, and
  J.~Zhao.
\newblock Untangling the global coronal magnetic field with multiwavelength
  observations.
\newblock 2020.

\bibitem[{Gudiksen} and {Nordlund}(2005)]{Gudiksen2005}
B.~V. {Gudiksen} and {\r{A}}.~{Nordlund}.
\newblock {An Ab Initio Approach to the Solar Coronal Heating Problem}.
\newblock \emph{\apj}, 618\penalty0 (2):\penalty0 1020--1030, Jan. 2005.
\newblock \doi{10.1086/426063}.

\bibitem[{Gudiksen} et~al.(2011){Gudiksen}, {Carlsson}, {Hansteen}, {Hayek},
  {Leenaarts}, and {Mart{\'\i}nez-Sykora}]{Gudiksen2011}
B.~V. {Gudiksen}, M.~{Carlsson}, V.~H. {Hansteen}, W.~{Hayek}, J.~{Leenaarts},
  and J.~{Mart{\'\i}nez-Sykora}.
\newblock {The stellar atmosphere simulation code Bifrost. Code description and
  validation}.
\newblock \emph{\aap}, 531:\penalty0 A154, July 2011.
\newblock \doi{10.1051/0004-6361/201116520}.

\bibitem[{Hansteen} et~al.(2017){Hansteen}, {Archontis}, {Pereira}, {Carlsson},
  {Rouppe van der Voort}, and {Leenaarts}]{Hansteen2017}
V.~H. {Hansteen}, V.~{Archontis}, T.~M.~D. {Pereira}, M.~{Carlsson}, L.~{Rouppe
  van der Voort}, and J.~{Leenaarts}.
\newblock {Bombs and Flares at the Surface and Lower Atmosphere of the Sun}.
\newblock \emph{\apj}, 839\penalty0 (1):\penalty0 22, Apr. 2017.
\newblock \doi{10.3847/1538-4357/aa6844}.

\bibitem[{Hayashi} et~al.(2018){Hayashi}, {Feng}, {Xiong}, and
  {Jiang}]{Hayashi2018}
K.~{Hayashi}, X.~{Feng}, M.~{Xiong}, and C.~{Jiang}.
\newblock {An MHD Simulation of Solar Active Region 11158 Driven with a
  Time-dependent Electric Field Determined from HMI Vector Magnetic Field
  Measurement Data}.
\newblock \emph{\apj}, 855\penalty0 (1):\penalty0 11, Mar. 2018.
\newblock \doi{10.3847/1538-4357/aaacd8}.

\bibitem[{Hewitt} et~al.(2014){Hewitt}, {Shelyag}, {Mathioudakis}, and
  {Keenan}]{Hewitt2014}
R.~L. {Hewitt}, S.~{Shelyag}, M.~{Mathioudakis}, and F.~P. {Keenan}.
\newblock {Plasma properties and Stokes profiles during the lifetime of a
  photospheric magnetic bright point}.
\newblock \emph{\aap}, 565:\penalty0 A84, May 2014.
\newblock \doi{10.1051/0004-6361/201322882}.

\bibitem[{Hoeksema} et~al.(2020){Hoeksema}, {Abbett}, {Bercik}, {Cheung},
  {DeRosa}, {Fisher}, {Hayashi}, {Kazachenko}, {Liu}, {Lumme}, {Lynch}, {Sun},
  and {Welsch}]{Hoeksema2020}
J.~T. {Hoeksema}, W.~P. {Abbett}, D.~J. {Bercik}, M.~C.~M. {Cheung}, M.~L.
  {DeRosa}, G.~H. {Fisher}, K.~{Hayashi}, M.~D. {Kazachenko}, Y.~{Liu},
  E.~{Lumme}, B.~J. {Lynch}, X.~{Sun}, and B.~T. {Welsch}.
\newblock {The Coronal Global Evolutionary Model: Using HMI Vector Magnetogram
  and Doppler Data to Determine Coronal Magnetic Field Evolution}.
\newblock \emph{arXiv e-prints}, art. arXiv:2006.14579, June 2020.

\bibitem[{Khomenko} et~al.(2020){Khomenko}, {Collados}, {Vitas}, and
  {Gonzalez-Morales}]{Khomenko2020}
E.~{Khomenko}, M.~{Collados}, N.~{Vitas}, and P.~A. {Gonzalez-Morales}.
\newblock {Influence of ambipolar and Hall effects on vorticity in 3D
  simulations of magneto-convection}.
\newblock \emph{arXiv e-prints}, art. arXiv:2009.09753, Sept. 2020.

\bibitem[{Kitiashvili} et~al.(2009){Kitiashvili}, {Kosovichev}, {Wray}, and
  {Mansour}]{Kitiashvili2009}
I.~N. {Kitiashvili}, A.~G. {Kosovichev}, A.~A. {Wray}, and N.~N. {Mansour}.
\newblock {Traveling Waves of Magnetoconvection and the Origin of the Evershed
  Effect in Sunspots}.
\newblock \emph{\apjl}, 700\penalty0 (2):\penalty0 L178--L181, Aug. 2009.
\newblock \doi{10.1088/0004-637X/700/2/L178}.

\bibitem[{Kitiashvili} et~al.(2010){Kitiashvili}, {Kosovichev}, {Wray}, and
  {Mansour}]{Kitiashvili2010}
I.~N. {Kitiashvili}, A.~G. {Kosovichev}, A.~A. {Wray}, and N.~N. {Mansour}.
\newblock {Mechanism of Spontaneous Formation of Stable Magnetic Structures on
  the Sun}.
\newblock \emph{\apj}, 719\penalty0 (1):\penalty0 307--312, Aug. 2010.
\newblock \doi{10.1088/0004-637X/719/1/307}.

\bibitem[{Kitiashvili} et~al.(2013){Kitiashvili}, {Kosovichev}, {Lele},
  {Mansour}, and {Wray}]{Kitiashvili2013}
I.~N. {Kitiashvili}, A.~G. {Kosovichev}, S.~K. {Lele}, N.~N. {Mansour}, and
  A.~A. {Wray}.
\newblock {Ubiquitous Solar Eruptions Driven by Magnetized Vortex Tubes}.
\newblock \emph{\apj}, 770\penalty0 (1):\penalty0 37, June 2013.
\newblock \doi{10.1088/0004-637X/770/1/37}.

\bibitem[{Kitiashvili} et~al.(2015){Kitiashvili}, {Kosovichev}, {Mansour}, and
  {Wray}]{Kitiashvili2015}
I.~N. {Kitiashvili}, A.~G. {Kosovichev}, N.~N. {Mansour}, and A.~A. {Wray}.
\newblock {Realistic Modeling of Local Dynamo Processes on the Sun}.
\newblock \emph{\apj}, 809\penalty0 (1):\penalty0 84, Aug. 2015.
\newblock \doi{10.1088/0004-637X/809/1/84}.

\bibitem[{Kitiashvili} et~al.(2020){Kitiashvili}, {Wray}, {Sadykov},
  {Kosovichev}, and {Mansour}]{Kitiashvili2020}
I.~N. {Kitiashvili}, A.~A. {Wray}, V.~{Sadykov}, A.~G. {Kosovichev}, and N.~N.
  {Mansour}.
\newblock {Realistic 3D MHD modeling of self-organized magnetic structuring of
  the solar corona}.
\newblock \emph{IAU Symposium}, 354:\penalty0 346--350, Jan. 2020.
\newblock \doi{10.1017/S1743921320001532}.

\bibitem[{Leenaarts} et~al.(2009){Leenaarts}, {Carlsson}, {Hansteen}, and
  {Rouppe van der Voort}]{Leenaarts2009}
J.~{Leenaarts}, M.~{Carlsson}, V.~{Hansteen}, and L.~{Rouppe van der Voort}.
\newblock {Three-Dimensional Non-LTE Radiative Transfer Computation of the CA
  8542 Infrared Line From a Radiation-MHD Simulation}.
\newblock \emph{\apjl}, 694\penalty0 (2):\penalty0 L128--L131, Apr. 2009.
\newblock \doi{10.1088/0004-637X/694/2/L128}.

\bibitem[{Leenaarts} et~al.(2013){Leenaarts}, {Pereira}, {Carlsson},
  {Uitenbroek}, and {De Pontieu}]{Leenaarts2013a}
J.~{Leenaarts}, T.~M.~D. {Pereira}, M.~{Carlsson}, H.~{Uitenbroek}, and B.~{De
  Pontieu}.
\newblock {The Formation of IRIS Diagnostics. I. A Quintessential Model Atom of
  Mg II and General Formation Properties of the Mg II h\&amp;k Lines}.
\newblock \emph{\apj}, 772\penalty0 (2):\penalty0 89, Aug. 2013.
\newblock \doi{10.1088/0004-637X/772/2/89}.

\bibitem[{Mart{\'\i}nez-Sykora} et~al.(2012){Mart{\'\i}nez-Sykora}, {De
  Pontieu}, and {Hansteen}]{Martinez-Sykora2012}
J.~{Mart{\'\i}nez-Sykora}, B.~{De Pontieu}, and V.~{Hansteen}.
\newblock {Two-dimensional Radiative Magnetohydrodynamic Simulations of the
  Importance of Partial Ionization in the Chromosphere}.
\newblock \emph{\apj}, 753\penalty0 (2):\penalty0 161, July 2012.
\newblock \doi{10.1088/0004-637X/753/2/161}.

\bibitem[{M{\"u}ller} et~al.(2020){M{\"u}ller}, {St. Cyr}, {Zouganelis},
  {Gilbert}, {Marsden}, {Nieves-Chinchilla}, {Antonucci}, {Auch{\`e}re},
  {Berghmans}, {Horbury}, {Howard}, {Krucker}, {Maksimovic}, {Owen}, {Rochus},
  {Rodriguez-Pacheco}, {Romoli}, {Solanki}, {Bruno}, {Carlsson}, {Fludra},
  {Harra}, {Hassler}, {Livi}, {Louarn}, {Peter}, {Sch{\"u}hle}, {Teriaca}, {del
  Toro Iniesta}, {Wimmer-Schweingruber}, {Marsch}, {Velli}, {De Groof},
  {Walsh}, and {Williams}]{Muller2020}
D.~{M{\"u}ller}, O.~C. {St. Cyr}, I.~{Zouganelis}, H.~R. {Gilbert},
  R.~{Marsden}, T.~{Nieves-Chinchilla}, E.~{Antonucci}, F.~{Auch{\`e}re},
  D.~{Berghmans}, T.~S. {Horbury}, R.~A. {Howard}, S.~{Krucker},
  M.~{Maksimovic}, C.~J. {Owen}, P.~{Rochus}, J.~{Rodriguez-Pacheco},
  M.~{Romoli}, S.~K. {Solanki}, R.~{Bruno}, M.~{Carlsson}, A.~{Fludra},
  L.~{Harra}, D.~M. {Hassler}, S.~{Livi}, P.~{Louarn}, H.~{Peter},
  U.~{Sch{\"u}hle}, L.~{Teriaca}, J.~C. {del Toro Iniesta}, R.~F.
  {Wimmer-Schweingruber}, E.~{Marsch}, M.~{Velli}, A.~{De Groof}, A.~{Walsh},
  and D.~{Williams}.
\newblock {The Solar Orbiter mission. Science overview}.
\newblock \emph{\aap}, 642:\penalty0 A1, Oct. 2020.
\newblock \doi{10.1051/0004-6361/202038467}.

\bibitem[Nita et~al.(2015)Nita, Fleishman, Kuznetsov, Kontar, and
  Gary]{Nita2015}
G.~Nita, G.~Fleishman, A.~Kuznetsov, E.~Kontar, and D.~Gary.
\newblock Three-dimensional radio and x-ray modeling and data analysis
  software: Revealing flare complexity.
\newblock \emph{Astrophysical Journal}, 799\penalty0 (2), 2015.
\newblock cited By 45.

\bibitem[Nita et~al.(2018)Nita, Viall, Klimchuk, Loukitcheva, Gary, Kuznetsov,
  and Fleishman]{Nita2018}
G.~Nita, N.~Viall, J.~Klimchuk, M.~Loukitcheva, D.~Gary, A.~Kuznetsov, and
  G.~Fleishman.
\newblock Dressing the coronal magnetic extrapolations of active regions with a
  parameterized thermal structure.
\newblock \emph{Astrophysical Journal}, 853\penalty0 (1), 2018.

\bibitem[{Nordlund} and {Stein}(1989)]{Nordlund1989}
{\r{A}}.~{Nordlund} and R.~F. {Stein}.
\newblock {Simulating Magnetoconvection}.
\newblock In R.~J. {Rutten} and G.~{Severino}, editors, \emph{NATO Advanced
  Science Institutes (ASI) Series C}, volume 263 of \emph{NATO Advanced Science
  Institutes (ASI) Series C}, page 453, Jan. 1989.

\bibitem[{Pereira} and {Uitenbroek}(2015)]{Pereira2015}
T.~M.~D. {Pereira} and H.~{Uitenbroek}.
\newblock {RH 1.5D: a massively parallel code for multi-level radiative
  transfer with partial frequency redistribution and Zeeman polarisation}.
\newblock \emph{\aap}, 574:\penalty0 A3, Feb. 2015.
\newblock \doi{10.1051/0004-6361/201424785}.

\bibitem[{Rast} et~al.(2020){Rast}, {Bello Gonz{\'a}lez}, {Bellot Rubio},
  {Cao}, {Cauzzi}, {DeLuca}, {De Pontieu}, {Fletcher}, {Gibson}, {Judge},
  {Katsukawa}, {Kazachenko}, {Khomenko}, {Landi}, {Mart{\'\i}nez Pillet},
  {Petrie}, {Qiu}, {Rachmeler}, {Rempel}, {Schmidt}, {Scullion}, {Sun},
  {Welsch}, {Andretta}, {Antolin}, {Ayres}, {Balasubramaniam}, {Ballai},
  {Berger}, {Bradshaw}, {Carlsson}, {Casini}, {Centeno}, {Cranmer}, {DeForest},
  {Deng}, {Erd{\'e}lyi}, {Fedun}, {Fischer}, {Gonz{\'a}lez Manrique}, {Hahn},
  {Harra}, {Henriques}, {Hurlburt}, {Jaeggli}, {Jafarzadeh}, {Jain},
  {Jefferies}, {Keys}, {Kowalski}, {Kuckein}, {Kuhn}, {Liu}, {Liu}, {Longcope},
  {McAteer}, {McIntosh}, {McKenzie}, {Miralles}, {Morton}, {Muglach}, {Nelson},
  {Panesar}, {Parenti}, {Parnell}, {Poduval}, {Reardon}, {Reep}, {Schad},
  {Schmit}, {Sharma}, {Socas-Navarro}, {Srivastava}, {Sterling}, {Suematsu},
  {Tarr}, {Tiwari}, {Tritschler}, {Verth}, {Vourlidas}, {Wang}, {Wang}, {NSO},
  {project}, {instrument scientists}, {the DKIST Science Working Group}, and
  {DKIST Critical Science Plan Community}]{Rast2020}
M.~P. {Rast}, N.~{Bello Gonz{\'a}lez}, L.~{Bellot Rubio}, W.~{Cao},
  G.~{Cauzzi}, E.~{DeLuca}, B.~{De Pontieu}, L.~{Fletcher}, S.~E. {Gibson},
  P.~G. {Judge}, Y.~{Katsukawa}, M.~D. {Kazachenko}, E.~{Khomenko}, E.~{Landi},
  V.~{Mart{\'\i}nez Pillet}, G.~J.~D. {Petrie}, J.~{Qiu}, L.~A. {Rachmeler},
  M.~{Rempel}, W.~{Schmidt}, E.~{Scullion}, X.~{Sun}, B.~T. {Welsch},
  V.~{Andretta}, P.~{Antolin}, T.~R. {Ayres}, K.~S. {Balasubramaniam},
  I.~{Ballai}, T.~E. {Berger}, S.~J. {Bradshaw}, M.~{Carlsson}, R.~{Casini},
  R.~{Centeno}, S.~R. {Cranmer}, C.~{DeForest}, Y.~{Deng}, R.~{Erd{\'e}lyi},
  V.~{Fedun}, C.~E. {Fischer}, S.~J. {Gonz{\'a}lez Manrique}, M.~{Hahn},
  L.~{Harra}, V.~M.~J. {Henriques}, N.~E. {Hurlburt}, S.~{Jaeggli},
  S.~{Jafarzadeh}, R.~{Jain}, S.~M. {Jefferies}, P.~H. {Keys}, A.~F.
  {Kowalski}, C.~{Kuckein}, J.~R. {Kuhn}, J.~{Liu}, W.~{Liu}, D.~{Longcope},
  R.~T.~J. {McAteer}, S.~W. {McIntosh}, D.~E. {McKenzie}, M.~P. {Miralles},
  R.~J. {Morton}, K.~{Muglach}, C.~J. {Nelson}, N.~K. {Panesar}, S.~{Parenti},
  C.~E. {Parnell}, B.~{Poduval}, K.~P. {Reardon}, J.~W. {Reep}, T.~A. {Schad},
  D.~{Schmit}, R.~{Sharma}, H.~{Socas-Navarro}, A.~K. {Srivastava}, A.~C.
  {Sterling}, Y.~{Suematsu}, L.~A. {Tarr}, S.~{Tiwari}, A.~{Tritschler},
  G.~{Verth}, A.~{Vourlidas}, H.~{Wang}, Y.-M. {Wang}, {NSO}, D.~{project},
  D.~{instrument scientists}, {the DKIST Science Working Group}, and t.~{DKIST
  Critical Science Plan Community}.
\newblock {Critical Science Plan for the Daniel K. Inouye Solar Telescope
  (DKIST)}.
\newblock \emph{arXiv e-prints}, art. arXiv:2008.08203, Aug. 2020.

\bibitem[{Rempel}(2014)]{Rempel2014}
M.~{Rempel}.
\newblock {Numerical Simulations of Quiet Sun Magnetism: On the Contribution
  from a Small-scale Dynamo}.
\newblock \emph{\apj}, 789\penalty0 (2):\penalty0 132, July 2014.
\newblock \doi{10.1088/0004-637X/789/2/132}.

\bibitem[{Rempel}(2017)]{Rempel2017}
M.~{Rempel}.
\newblock {Extension of the MURaM Radiative MHD Code for Coronal Simulations}.
\newblock \emph{\apj}, 834\penalty0 (1):\penalty0 10, Jan. 2017.
\newblock \doi{10.3847/1538-4357/834/1/10}.

\bibitem[{Rempel} et~al.(2009{\natexlab{a}}){Rempel}, {Sch{\"u}ssler},
  {Cameron}, and {Kn{\"o}lker}]{Rempel2009a}
M.~{Rempel}, M.~{Sch{\"u}ssler}, R.~H. {Cameron}, and M.~{Kn{\"o}lker}.
\newblock {Penumbral Structure and Outflows in Simulated Sunspots}.
\newblock \emph{Science}, 325\penalty0 (5937):\penalty0 171, July
  2009{\natexlab{a}}.
\newblock \doi{10.1126/science.1173798}.

\bibitem[{Rempel} et~al.(2009{\natexlab{b}}){Rempel}, {Sch{\"u}ssler}, and
  {Kn{\"o}lker}]{Rempel2009}
M.~{Rempel}, M.~{Sch{\"u}ssler}, and M.~{Kn{\"o}lker}.
\newblock {Radiative Magnetohydrodynamic Simulation of Sunspot Structure}.
\newblock \emph{\apj}, 691\penalty0 (1):\penalty0 640--649, Jan.
  2009{\natexlab{b}}.
\newblock \doi{10.1088/0004-637X/691/1/640}.

\bibitem[{Rimmele} et~al.(2020){Rimmele}, {Warner}, and {Keil}]{rimmele2020}
R.~{Rimmele}, T, M.~{Warner}, and e.~a. {Keil}, S.~L.
\newblock {The Daniel K. Inouye Solar Telescope – Observatory Overview}.
\newblock \emph{Solar Physics}, 295\penalty0 (172):\penalty0 162, Dec. 2020.
\newblock \doi{10.1007/s11207-020-01736-7}.

\bibitem[{Sainz Dalda} et~al.(2019){Sainz Dalda}, {de la Cruz Rodr{\'\i}guez},
  {De Pontieu}, and {Go{\v{s}}i{\'c}}]{SainzDalda2019}
A.~{Sainz Dalda}, J.~{de la Cruz Rodr{\'\i}guez}, B.~{De Pontieu}, and
  M.~{Go{\v{s}}i{\'c}}.
\newblock {Recovering Thermodynamics from Spectral Profiles observed by IRIS: A
  Machine and Deep Learning Approach}.
\newblock \emph{\apjl}, 875\penalty0 (2):\penalty0 L18, Apr. 2019.
\newblock \doi{10.3847/2041-8213/ab15d9}.

\bibitem[{Sch{\"u}ssler} and {V{\"o}gler}(2006)]{Schuessler2006}
M.~{Sch{\"u}ssler} and A.~{V{\"o}gler}.
\newblock {Magnetoconvection in a Sunspot Umbra}.
\newblock \emph{\apjl}, 641\penalty0 (1):\penalty0 L73--L76, Apr. 2006.
\newblock \doi{10.1086/503772}.

\bibitem[{Siu-Tapia} et~al.(2018){Siu-Tapia}, {Rempel}, {Lagg}, and
  {Solanki}]{SiuTapia2018}
A.~L. {Siu-Tapia}, M.~{Rempel}, A.~{Lagg}, and S.~K. {Solanki}.
\newblock {Evershed and Counter-Evershed Flows in Sunspot MHD Simulations}.
\newblock \emph{\apj}, 852\penalty0 (2):\penalty0 66, Jan. 2018.
\newblock \doi{10.3847/1538-4357/aaa007}.

\bibitem[{Skartlien} et~al.(2000){Skartlien}, {Stein}, and
  {Nordlund}]{Skartlien2000}
R.~{Skartlien}, R.~F. {Stein}, and {\r{A}}.~{Nordlund}.
\newblock {Excitation of Chromospheric Wave Transients by Collapsing Granules}.
\newblock \emph{\apj}, 541\penalty0 (1):\penalty0 468--488, Sept. 2000.
\newblock \doi{10.1086/309414}.

\bibitem[{Socas-Navarro} et~al.(2015){Socas-Navarro}, {de la Cruz
  Rodr{\'\i}guez}, {Asensio Ramos}, {Trujillo Bueno}, and {Ruiz
  Cobo}]{SocasNavarro2015}
H.~{Socas-Navarro}, J.~{de la Cruz Rodr{\'\i}guez}, A.~{Asensio Ramos},
  J.~{Trujillo Bueno}, and B.~{Ruiz Cobo}.
\newblock {An open-source, massively parallel code for non-LTE synthesis and
  inversion of spectral lines and Zeeman-induced Stokes profiles}.
\newblock \emph{\aap}, 577:\penalty0 A7, May 2015.
\newblock \doi{10.1051/0004-6361/201424860}.

\bibitem[{Stein} and {Nordlund}(2001)]{Stein2001}
R.~F. {Stein} and {\r{A}}.~{Nordlund}.
\newblock {Solar Oscillations and Convection. II. Excitation of Radial
  Oscillations}.
\newblock \emph{\apj}, 546\penalty0 (1):\penalty0 585--603, Jan. 2001.
\newblock \doi{10.1086/318218}.

\bibitem[{Stein} and {Nordlund}(2012)]{Stein2012}
R.~F. {Stein} and {\r{A}}.~{Nordlund}.
\newblock {On the Formation of Active Regions}.
\newblock \emph{\apjl}, 753\penalty0 (1):\penalty0 L13, July 2012.
\newblock \doi{10.1088/2041-8205/753/1/L13}.

\bibitem[{Tomczyk} et~al.(2016){Tomczyk}, {Landi}, {Burkepile}, {Casini},
  {DeLuca}, {Fan}, {Gibson}, {Lin}, {McIntosh}, {Solomon}, {Toma}, {Wijn}, and
  {Zhang}]{Tomczyk}
S.~{Tomczyk}, E.~{Landi}, J.~T. {Burkepile}, R.~{Casini}, E.~E. {DeLuca},
  Y.~{Fan}, S.~E. {Gibson}, H.~{Lin}, S.~W. {McIntosh}, S.~C. {Solomon},
  G.~{Toma}, A.~G. {Wijn}, and J.~{Zhang}.
\newblock {Scientific objectives and capabilities of the Coronal Solar
  Magnetism Observatory}.
\newblock \emph{Journal of Geophysical Research (Space Physics)}, 121\penalty0
  (8):\penalty0 7470--7487, Aug. 2016.
\newblock \doi{10.1002/2016JA022871}.

\bibitem[{Vitas} et~al.(2011){Vitas}, {Fischer}, {V{\"o}gler}, and
  {Keller}]{Vitas2011}
N.~{Vitas}, C.~E. {Fischer}, A.~{V{\"o}gler}, and C.~U. {Keller}.
\newblock {Fast horizontal flows in a quiet sun MHD simulation and their
  spectroscopic signatures}.
\newblock \emph{\aap}, 532:\penalty0 A110, Aug. 2011.
\newblock \doi{10.1051/0004-6361/201015773}.

\bibitem[{V{\"o}gler} and {Sch{\"u}ssler}(2007)]{Vogler2007}
A.~{V{\"o}gler} and M.~{Sch{\"u}ssler}.
\newblock {A solar surface dynamo}.
\newblock \emph{\aap}, 465\penalty0 (3):\penalty0 L43--L46, Apr. 2007.
\newblock \doi{10.1051/0004-6361:20077253}.

\bibitem[{V{\"o}gler} et~al.(2005){V{\"o}gler}, {Shelyag}, {Sch{\"u}ssler},
  {Cattaneo}, {Emonet}, and {Linde}]{Vogler2005}
A.~{V{\"o}gler}, S.~{Shelyag}, M.~{Sch{\"u}ssler}, F.~{Cattaneo}, T.~{Emonet},
  and T.~{Linde}.
\newblock {Simulations of magneto-convection in the solar photosphere.
  Equations, methods, and results of the MURaM code}.
\newblock \emph{\aap}, 429:\penalty0 335--351, Jan. 2005.
\newblock \doi{10.1051/0004-6361:20041507}.

\bibitem[{{\v{S}}t{\v{e}}p{\'a}n} et~al.(2015){{\v{S}}t{\v{e}}p{\'a}n},
  {Trujillo Bueno}, {Leenaarts}, and {Carlsson}]{Stepan2015}
J.~{{\v{S}}t{\v{e}}p{\'a}n}, J.~{Trujillo Bueno}, J.~{Leenaarts}, and
  M.~{Carlsson}.
\newblock {Three-dimensional Radiative Transfer Simulations of the Scattering
  Polarization of the Hydrogen Ly{\ensuremath{\alpha}} Line in a
  Magnetohydrodynamic Model of the Chromosphere-Corona Transition Region}.
\newblock \emph{\apj}, 803\penalty0 (2):\penalty0 65, Apr. 2015.
\newblock \doi{10.1088/0004-637X/803/2/65}.

\bibitem[{Wray} et~al.(2018){Wray}, {Bensassiy}, {Kitiashvili}, {Mansour}, and
  {Kosovichev}]{Wray2018}
A.~A. {Wray}, K.~{Bensassiy}, I.~N. {Kitiashvili}, N.~N. {Mansour}, and A.~G.
  {Kosovichev}.
\newblock \emph{{Realistic Simulations of Stellar Radiative MHD}}, page~39.
\newblock 2018.

\bibitem[Wright et~al.(2019)Wright, Cheung, Thomas, Galvez, Szenicer, Jin,
  Muñoz-Jaramillo, and Fouhey]{Wright2019}
P.~J. Wright, M.~C.~M. Cheung, R.~Thomas, R.~Galvez, A.~Szenicer, M.~Jin,
  A.~Muñoz-Jaramillo, and D.~Fouhey.
\newblock {DeepEM: Demonstrating a Deep Learning Approach to DEM Inversion},
  Mar. 2019.
\newblock URL \url{https://doi.org/10.5281/zenodo.2587015}.

\bibitem[{Yang} et~al.(2020a){Yang}, {Bethge}, {Tian}, {Tomczyk}, {Morton},
  {Del Zanna}, {McIntosh}, {Karak}, {Gibson}, {Samanta}, {He}, {Chen}, and
  {Wang}]{Yang2020a}
Z.~{Yang}, C.~{Bethge}, H.~{Tian}, S.~{Tomczyk}, R.~{Morton}, G.~{Del Zanna},
  S.~W. {McIntosh}, B.~B. {Karak}, S.~{Gibson}, T.~{Samanta}, J.~{He},
  Y.~{Chen}, and L.~{Wang}.
\newblock {Global maps of the magnetic field in the solar corona}.
\newblock \emph{Science}, 369\penalty0 (6504):\penalty0 694--697, Aug. 2020a.
\newblock \doi{10.1126/science.abb4462}.

\bibitem[Yang et~al.(2020b)Yang, Tian, Tomczyk, Morton, Bai, Samanta, and
  Chen]{Yang2020b}
Z.~Yang, H.~Tian, S.~Tomczyk, R.~Morton, X.~Bai, T.~Samanta, and Y.~Chen.
\newblock Mapping the magnetic field in the solar corona through
  magnetoseismology.
\newblock \emph{Science China Technological Sciences}, 63\penalty0
  (11):\penalty0 2357--2368, Nov 2020b.
\newblock ISSN 1869-1900.
\newblock \doi{10.1007/s11431-020-1706-9}.
\newblock URL \url{https://doi.org/10.1007/s11431-020-1706-9}.

\end{thebibliography}

\end{document}